\documentclass{article}
\usepackage[twocolumn]{emulateapj}
\submitted{To appear in the Astrophysical Journal 1999 March 10 issue}

\newcommand\etal{{ et al. }}
\def\lsim{\mathrel{\rlap{\lower 4pt \hbox{\hskip 1pt $\sim$}}\raise 1pt \hbox
        {$<$}}}
\def\gsim{\mathrel{\rlap{\lower 4pt \hbox{\hskip 1pt $\sim$}}\raise 1pt \hbox
        {$>$}}}
 
\lefthead{}
\righthead{Evolution of 3-9 $M_\odot$ Stars from Z=0.001 to 0.03\\
and metallicity effects on Type Ia Supernovae}

\begin{document}
\title{Evolution of 3 - 9 $M_\odot$ Stars for Z = 0.001 -- 0.03\\
and Metallicity Effects on Type Ia Supernovae}

\bigskip

\author{HIDEYUKI UMEDA, KEN'ICHI NOMOTO}
\affil{Research Center for the Early Universe, and \\ Department of
Astronomy, School of Science, University of Tokyo, \\ Bunkyo-ku, 
Tokyo 113-0033, Japan \\E-mail: umeda@iron.astron.s.u-tokyo.ac.jp; 
nomoto@astron.s.u-tokyo.ac.jp}

\author{HITOSHI YAMAOKA, } 
\affil{ Department of Physics, Faculty of Science, 
 Kyushu University, \\ Ropponmatsu, Fukuoka 810-8560, Japan
 \\E-mail: yamaoka@rc.kyushu-u.ac.jp}

\author{SHINYA WANAJO}
\affil{ Division of Theoretical Astrophysics \\
 National Astronomical Observatory,  Mitaka, Tokyo 181-8588, Japan   
 \\E-mail: wanajo@diamond.mtk.nao.ac.jp}

\begin{abstract}
 Recent observations have revealed that Type Ia supernovae
(SNe Ia) are not perfect standard
candles but show some variations in their absolute magnitudes,
light curve shapes, and spectra. The C/O ratio in the SNe Ia progenitors
(C-O white dwarfs) may be related to this variation. 
In this work, we systematically investigate the effects of
stellar mass ($M$) and metallicity ($Z$) on the C/O ratio and 
its distribution in the C-O white dwarfs
by calculating stellar evolution from the main-sequence through the 
end of the second dredge-up for $M=3-9 M_\odot$ and $Z=0.001-0.03$.
We find that the total 
carbon mass fraction just before SN Ia explosion varies in
the range 0.36 -- 0.5.
We also calculate the metallicity dependence of 
the main-sequence-mass range of the SN Ia progenitor 
white dwarfs.  Our results show that the maximum main-sequence
mass to form C-O white dwarfs decreases significantly
toward lower metallicity, and the number of SN Ia progenitors
may be underestimated if metallicity effect is neglected.
We discuss the implications of these results on the variation
of SNe Ia, determination of cosmological 
parameters, luminosity function of
white dwarfs, and the galactic chemical evolution.
\end{abstract}


\section{Introduction}

 Type Ia supernovae (SNe Ia) are characterized by the absence of hydrogen and
the presence of strong Si II line in the early time
spectra. Majority of SNe Ia have
almost the same absolute brightness, thereby being good distance 
indicators for determining cosmological parameters 
(e.g., Branch and Tammann 1992; Sandage and Tammann 1993;
 Nomoto, Iwamoto, \& Kishimoto 1997 for a recent review).  
However, 
SNe Ia are not perfect standard candles but show some variations in their 
absolute magnitudes, light curve shapes, and spectra
(Phillips 1993; Branch, Romanishin, \& Baron 1996; Hamuy \etal 1996). 
It is important to understand the source of these variations 
and the dependence on the metallicity in order to use high-redshift
SNe Ia for determining the cosmological parameters. 

 It is widely accepted that SNe Ia are thermonuclear
explosions of accreting C-O white dwarfs (WDs), although detailed explosion 
mechanism is still under debate (e.g. 
Ruiz-Lapuente, Canal, \& Isern 1997). 
Previous studies have shown that typical SNe Ia are well
described by the Chandrasekhar mass model, in which accreting C-O WDs
explode when the masses of these WDs approach the
critical mass $M_{\rm Ia} \simeq 1.37-1.38 M_\odot$
(Nomoto, Thielemann, \& Yokoi 1984; H\"oflich \& Khokhlov 1996). 
The explosion is induced
by the carbon ignition in the central region and the subsequent propagation
of the burning front. Behind the burning front, explosive
burning produces a large amount of $^{56}$Ni. 
There are two models for the burning front propagation: 
the deflagration model 
(e.g., Ivanova \etal 1974; Nomoto, Sugimoto, \& Neo 1976; 
Nomoto \etal 1984)    
and the delayed (or late) detonation (DD) model
(Khokhlov 1991; Yamaoka \etal 1992; Woosley \& Weaver 1994).

 There are several possible sources of the variations of 
brightness and spectra: 
for example, variations of the age of the WDs and the mass accretion
rates, which affect the ignition density, variations of the WD
progenitor's main-sequence mass and metallicity, which affect
the C/O ratio in the WDs, and the variation of 
the deflagration-detonation
transition density ($\rho_{tr}$) in the DD model.
These effects modify the explosion energy and the synthesized
$^{56}$Ni mass, which affects
 the maximum brightness and shape of the SN Ia light curve.

 Regarding the C/O ratio, previous explosion models have been constructed
by assuming a uniform mass fraction ratio C/O = 1, though previous
results indicated C/O $< 1$ in the central region (e.g. Mazzitelli, \&
D'Antona 1986 a,b; Schaller \etal 1992, SSMM hereafter). 
This is partly for simplicity and partly due to 
 its strong sensitivity on the treatment of
convection and the $^{12}$C$(\alpha,\gamma)^{16}$O rate.
However, the effect of the different C/O ratio is not negligible. 
For example,
recently H\"oflich, Wheeler, \& Thielemann (1998)
constructed explosion models by arbitrarily changing the C/O ratio
to C/O = 2/3
and found some significant effects on the light curve and spectra.
 
 Intermediate mass stars by definition
burn only hydrogen and helium, and form C-O cores after He burning.
These cores will become C-O WDs, thus becoming SNe Ia 
progenitors in certain binary systems. 
Previously, several groups
have investigated intermediate mass stars, but they have not
paid much attention to the C/O ratio and  the C-O core mass size
with its systematic dependence
on the stellar mass and metallicity, as well as its effects on SNe Ia.
For example, SSMM and a series of subsequent papers
(Schaerer \etal 1993; Charbonnel \etal 1993)
studied the influence of the metallicity on the
life times, luminosities and the chemical structure. However, they
only showed the central and surface chemical abundances and did not
show the C-O core mass sizes. Salaris \etal (1997) showed the radial
profile of the C/O ratio 
in the intermediate mass stars as well as the
C-O core mass size, but only for solar metallicity.
Vassiliadis \& Wood (1993) calculated C-O core mass sizes for 
the stars below 5 $M_\odot$, but did not pay much attention to the C/O ratio.
The works prior to 1992, such as Castellani \etal (1990), 
Tornamb\'e \& Chieffi (1986), Becker \& Iben (1979),
used Los Alamos (Huebner \etal 1977) or Cox \& Stewart (1970)
opacity tables, and hence their results would be significantly
different if latest OPAL opacity (Iglesias \& Rogers 1993) was used.

 In this paper, using updated input physics,
we evolve the intermediate mass ($3-9  M_\odot$)
stars and examine the stellar metallicity and mass
dependence of the C/O ratio, its radial profile,
and the C-O core mass size. 
These quantities are sensitive to
the treatment of convective mixing; however, for the same treatment
of convection, the  $^{12}$C$(\alpha,\gamma)^{16}$O rate dominantly
determines these. As described in section 2,
we calibrate this rate by the massive star nucleosynthesis.

 We also obtain metallicity dependence of zero-age main-sequence 
(ZAMS) mass ranges of SN Ia progenitors, and especially its upper
limit $M_{\rm UP}$, where  $M_{\rm UP}$ is the ZAMS mass 
at the border between the star that ignites carbon under
non-degenerate condition and the star that forms a strong degenerate
C-O core. This quantity
is important for various applications, such as galactic chemical evolution
and the luminosity function of WDs in our galaxy. 
However, earlier determination of $M_{\rm UP}$ was based on
old radiative opacity tables (e.g. Becker \& Iben 1979, BI79 hereafter;
Castellani \etal 1985, CCPT hereafter; Tornamb\'e \& Chieffi 1986;
Castellani \etal 1990) or not by exact calculations (Schaerer \etal 1993).
Also the calculations during the 1980s and 
early 1990s often assumed strong overshooting or breathing pulses, 
which may not be preferable with the use of the OPAL opacity.
Their results are qualitatively 
consistent with our results but quantitatively different.
 Our $M_{\rm UP}$ is significantly smaller than the
estimates in BI79 for all Z we calculated, and in Schaerer \etal (1993)
for Z$\lsim 0.01$, and it shows a large decrease toward 
lower metallicity. Such a dependence would be important for the
distributions of SNe Ia brightness as a function of metallicity.

 This paper is organized as follows. In the next section, our numerical
code, input physics and parameters are
described. Section 3 shows our results and compares with 
previous works. Their implications are discussed in Section 4.

\section{Numerical Code and Assumptions}

 We modify and update the Henyey type stellar evolution 
code (Saio \& Nomoto 1988).
Important improvements include the use of the 
OPAL radiative opacity (Iglesias \& Rogers 1993), 
neutrino emissivity (Itoh, Nishikawa, \& Kohyama 1996)
and a nuclear reaction network with larger number of isotopes.
Most of the nuclear reaction rates are provided by Thielemann (1995).
 This network is coupled to the dynamical
equations for stellar evolution
{\sl implicitly} in a similar manner as SSMM and Weaver \& Woosley (1993).
That is, the network with a reduced number of isotopes is run 
during Henyey relaxation. Nuclear energy generation rates between the
reduced and full networks are almost identical.
After dynamics is converged, nucleosynthesis is calculated with a full
network code for He burning. The isotopes included are shown in 
Table 1. For H burning, the network code of Saio \& Nomoto (1988) is used
with updated nuclear reaction rates.

\placetable{tab1}

 Treatment of convection is one of the serious uncertainties
in the theory of stellar evolution. We adopt the Schwarzschild
criterion for convective instability and the diffusive treatment 
for the convective mixing (Spruit 1992; Saio \& Nomoto 1998).
This formalism contains a parameter $f_k$ of order unity
 which is proportional to the diffusion coefficient, and also to the 
square root of the ratio of
``solute'' to thermal diffusivity. We found that the results were
not sensitive to this parameter at least in the range 
$f_k \simeq 0.1 - 1$. In this paper, we set $f_k = 0.3$.
We use the mixing length parameter $\alpha$ = 1.5 for the
outermost super-adiabatic convection.
Chieffi, Straniero, \& Salaris (1995) found that a larger
 value ($\alpha \simeq $2.25) fits observations well with the use of
OPAL opacity. However, as shown in the next section, in our code 
the evolutionary tracks in the H-R
diagram are closer to those of SSMM for $\alpha$ = 1.5.
Also as described in the next section, the core property, especially
the central C/O ratio, is insensitive to this parameter. Therefore, our
conclusions in this paper are little affected by the choice of
a different $\alpha$.
We neither artificially enhance nor suppress convective regions
from the  Schwarzschild criterion throughout the calculations, and hence
we do not include convective overshooting 
and breathing pulses at the end of He burning. It is considered that 
with the use of the OPAL opacity these effects are not so large
as considered before (e.g. Stothers \& Chin 1991); however,
whether these effects
are significant or not is still uncertain. 

 Once the treatment of convection is fixed, 
the $^{12}$C$(\alpha,\gamma)^{16}$O rate dominantly
determines the C/O ratio after He burning and the C-O core mass size.
Since the track in the H-R diagram is not sensitive to
the $^{12}$C$(\alpha,\gamma)^{16}$O rate, it is very difficult to constrain
this rate from the surface property of the stars, such as 
spectral observations of clusters. In this paper this rate
is chosen to be 1.5 times the value of Caughlan \& Fowler 
(1988, CF88 hereafter). This choice, denoted by 1.5$\times$ CF88, 
with the same treatment for the convection
yields the central carbon mass fraction X($^{12}$C) $\simeq 0.20$
after the He burning of the 25 M$_\odot$ star (Umeda \& Nomoto 1998).
Therefore, it approximately reproduces the central C/O ratio of 
the Nomoto \& Hashimoto (1988, NH88 hereafter) model, being expected 
to be consistent with solar abundances of Ne/O and Mg/O
in massive star yield (Hashimoto, Nomoto, \& Shigeyama 1989;
Thielemann, Nomoto, \& Hashimoto 1996; Woosley \& Weaver 1995). 
We note that although the $^{12}$C$(\alpha,\gamma)^{16}$O rate is
constrained most stringently by the nucleosynthesis argument of 
massive star yield (Weaver \& Woosley 1993;
Thielemann \etal 1996), our knowledge
of this rate is still limited, because there are still
uncertainties in the chemical evolution, massive star
evolution, and the explosive nuclear burning models.
In Section 3.3, we examine the dependence of this rate by applying
the rates of 1.7, 1.5, and 1.0 times the rate of CF88 and the rate
of Caughlan \etal (1985, CF85 hereafter).
More detailed investigation of the C/O ratio in massive stars,
and its dependence on $^{12}$C$(\alpha,\gamma)^{16}$O 
rate and treatments of convection will be given in Umeda \& Nomoto (1998).

 We adopt the solar chemical composition by Anders \& Grevesse
(1989), the solar He abundance $Y_\odot$ =0.27753
and metallicity  $Z_\odot$=0.02 (Bahcall \& Pinsonneault 1995).
We assume that the initial 
(zero age main sequence or ZAMS) helium abundance depends on
metallicity $Z$ as  
$ Y (Z)= Y_{p} + (\Delta Y/\Delta Z) Z$ (for $Z\leq 0.02)$,
where $\Delta Y/\Delta Z$ is a constant and $Y_p$ is the 
primordial helium abundance. In this paper we adopt  $Y_p
=0.247 $ (Tytler, Fan, \& Burles 1996), and thus $\Delta Y/\Delta Z = 1.5265$. 
 For $Z> 0.02$,  we only show the case of $(Z,Y)$=(0.03, 0.28). 
Different $Y$ yields different stellar evolution. Especially, the star is
brighter for a larger $Y$. However, as shown below the effects of
changing $Y$ is not very large.

 We also include an empirical mass loss rate (de Jager,
 Nieuwenhuijzen, \& van der Hucht 1988)
being scaled with metallicity as
$(Z/0.02)^{0.5}$ (Kudritzki \etal 1989).

\section{Results}

 We calculated the evolution of intermediate-mass ($3-9 M_\odot$)
stars for metallicity $Z$=0.001 -- 0.03. 
 Figure 1 shows the evolution in the
H-R diagram and the central density ($\rho_c$) vs.
central temperature ($T_c$) diagram
of our 5 $M_\odot$ ($Z$=0.02) model compared with SSMM. The 
luminosity of our model tends to be lower
than SSMM and $\rho_c$ is higher for the same $T_c$.
There are many differences in the inputs, which may be the reasons
for these differences: we adopt $Y=0.2775$ for $Z=0.02$, while SSMM
assumed $Y=0.3$. 
Our choice of the  $^{12}$C$(\alpha,\gamma)^{16}$O rate (1.5 $\times$
CF88) are smaller than CF85 $\sim$ 2.3 $\times$ CF88 in SSMM.
We use the larger nuclear reaction network. The treatment of
convection is different. Especially they assumed overshooting from 
a convective core, which we do not.
Figure 1 shows that the difference of $Y$
alone does not explain the discrepancy.
Also, we find that the tracks on the H-R diagrams are almost
identical even though the $^{12}$C$(\alpha,\gamma)^{16}$O rate is
different. Therefore,
probably the treatment of convection is mainly responsible for
the differences. Indeed, the tracks on the H-R diagrams of the
no-overshooting models in Bressan \etal (1993) are similar to our results.

\begin{figure}
\plotone{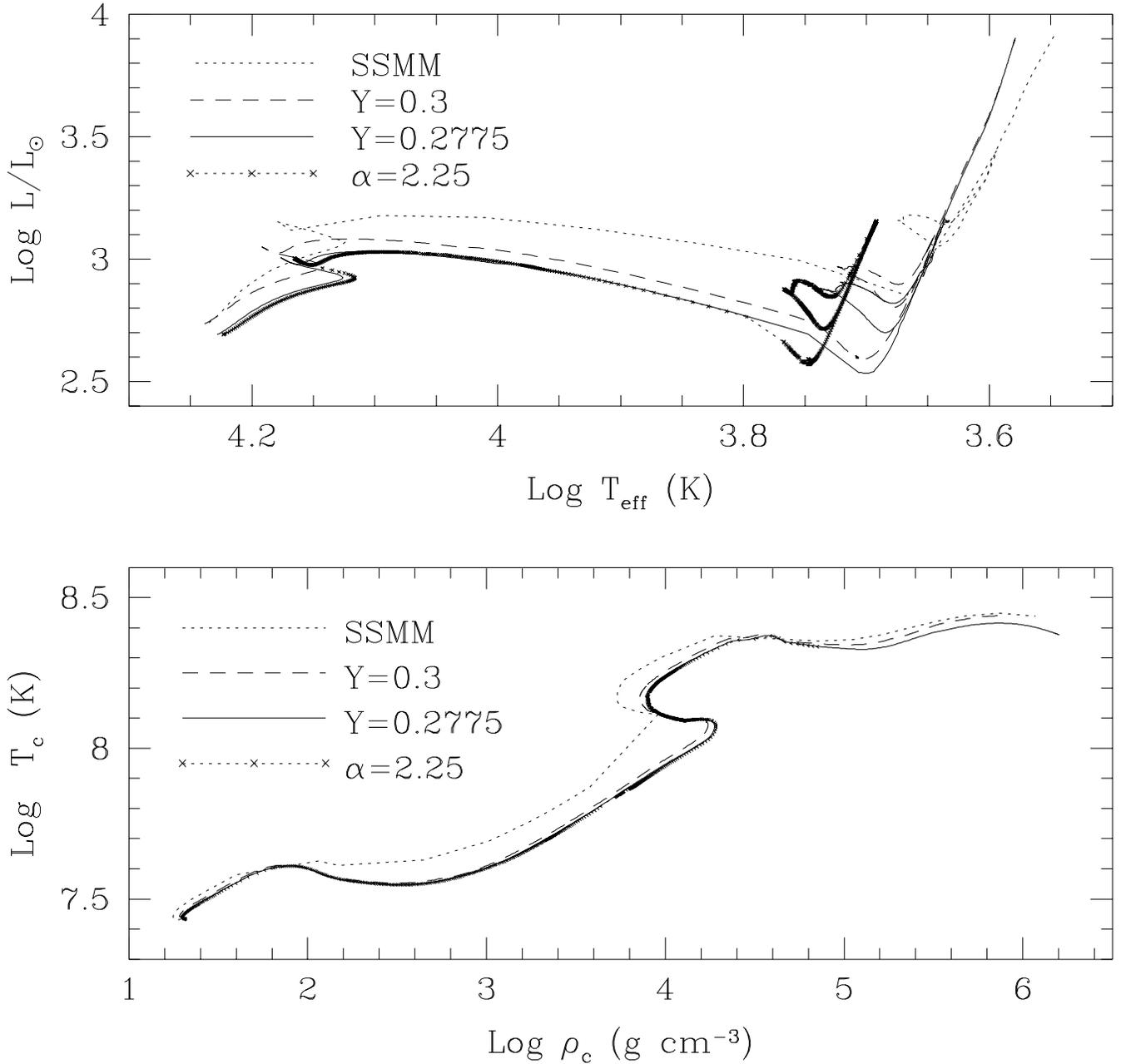}
\figcaption[coFIG1.epsi]{ Evolution in the
 H-R diagram (upper panel) and the central density ($\rho_c$) vs.
central temperature ($T_c$) diagram (lower panel) of our 5 $M_\odot$
($Z$=0.02) models for $Y=0.2775$ (solid), compared with SSMM
(dotted; Schaller \etal 1992). Dependence on $Y$ is small as
seen from the case of $Y=0.3$ (dashed). A model with a larger 
mixing length parameter $\alpha$ (=2.25) is shown in a dotted curve
with crosses.
\label{FIG1}}
\end{figure}

\placefigure{FIG1}

 In this figure we also show the effect of mixing length
parameter $\alpha$ for the outermost super-adiabatic convection. 
This parameter changes the location in the
H-R diagram after a star becomes a red giant. On the other hand, the
central $\rho_c$ -- $T_c$ track is almost independent of this
parameter. As a result, the central C/O ratio is almost identical
between the $\alpha$ = 1.5 and 2.25 cases. Therefore, the choice of this
parameter is not so important for the purpose of this paper.
Chieffi \etal (1995) found that
$\alpha$ = 2.25 fits observations well; however, the evolutionary tracks
in the H-R diagram are
closer to those of SSMM for $\alpha$ = 1.5 than 2.25, and hence
we adopt $\alpha$ = 1.5 in the followings.

\subsection {Metallicity effects on stellar evolution}

 In the ranges of stellar masses and $Z$ considered in this paper, the most 
important metallicity effect is that the radiative opacity is
smaller for lower $Z$. Therefore, a star with lower $Z$ is brighter,
thus having a shorter
lifetime than a star with the same mass but higher $Z$.
In this sense, the effect of reducing metallicity for these stars
is almost equivalent to increasing a stellar mass. This is verified
in Figure 2, which shows the H-R
diagrams of the 5$M_\odot$ models for different metallicities (upper panel)
and the models with $Z=0.02$ 
for different stellar masses (lower panel). 

\begin{figure}
\plotone{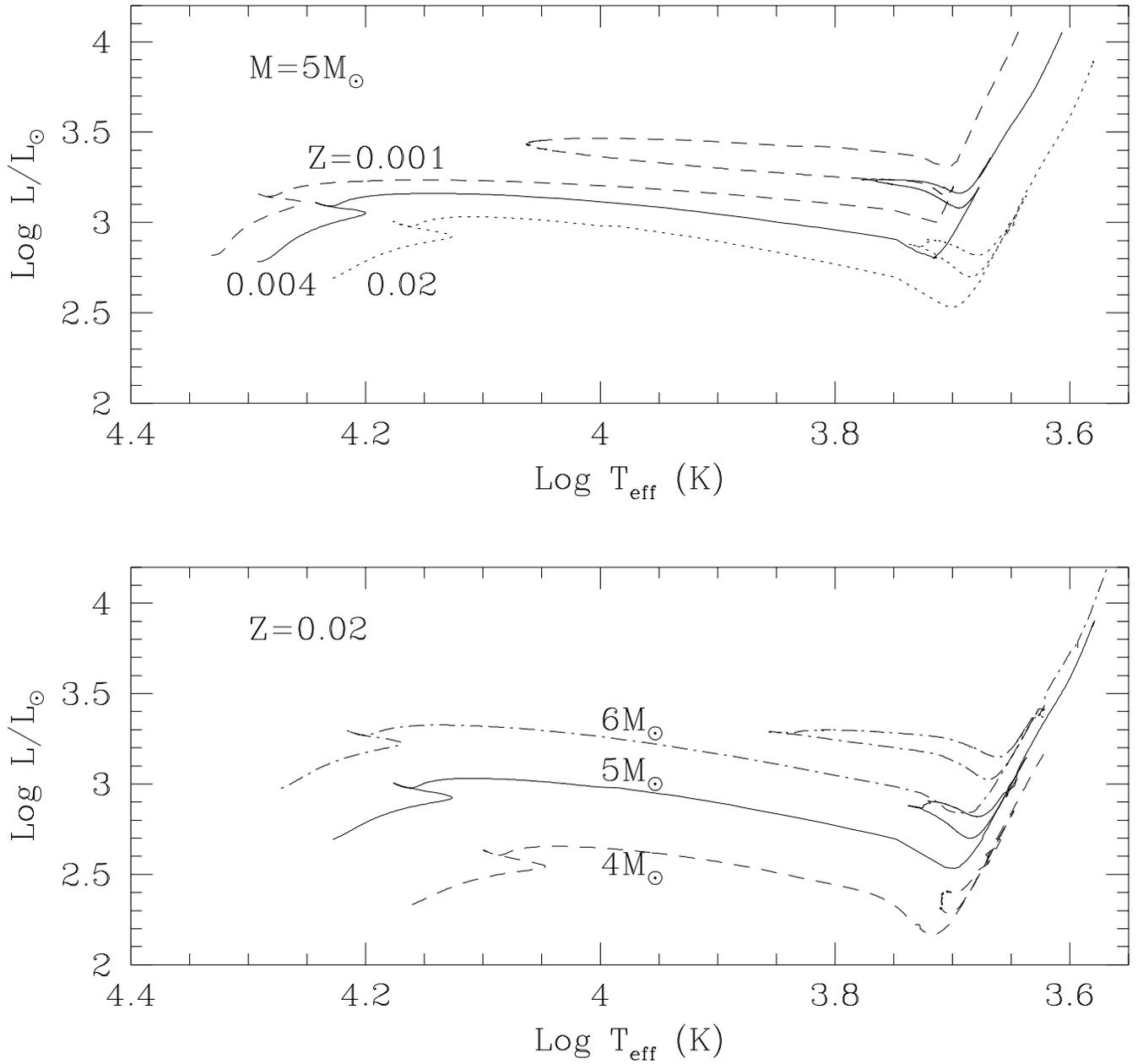}
\figcaption[coFIG2.epsi]{Evolution in the
 H-R diagrams of the 5$M_\odot$ models for several metallicities 
$Z$ (upper panel)
and the $Z=0.02$ models for several stellar masses (lower panel). 
\label{FIG2}}
\end{figure}

\placefigure{FIG2}

 For stars with larger masses and/or smaller $Z$, 
the luminosity is higher at the same evolutionary phase. With a higher
nuclear energy generation rate, these stars have larger convective
cores during H and He burning, thus forming larger
He and C-O cores. This property is depicted in Figure 3(a) and (b).
 
 Figure 3(a) shows abundances in mass fraction in 
the inner core of the 7 $M_\odot$ star for $Z$=0.001 at age $t$=48.45 Myr, 
with the central temperature and 
density log $T_c$ (K)=8.58 and log $\rho_c$ (g cm$^{-3}$) =6.82.
In Figure 3(a), the He burning shell is located around the mass coordinate
$M_r=1.1 M_\odot$, which
indicates that the C-O core of this model will exceed 1.1  $M_\odot$.
Since this is well above the critical mass for the off-center carbon
ignition, this star will ignite carbon in the non-degenerate
region and will not form a C-O WD but form an O-Ne-Mg core (Nomoto 1984).

 If the star is less massive or has higher metal abundance than 
the model in Figure 3(a), the C-O core can be smaller and the carbon
ignition is avoided as shown in Figure 3(b)
for the 7 $M_\odot$ model of Z=0.02 at age 
t=51.87 Myr (at the end of the second dredge-up).
The dredge-up is a phenomenon during which the surface convection is
penetrating inwards and the inner material is being brought up to the surface.
The first dredge-up
occurs before or around the He ignition and the second one 
occurs during the He shell burning stage for intermediate-mass stars. 
During the second dredge-up phase, the thick He shell
which is seen in Figure 3(a) becomes thinner and thinner because of 
the C-O core growth through He
shell burning and the penetration of the surface convection. Therefore,
as shown in Figure 3(b), the He shell becomes very thin 
(in mass coordinate) at the end of the second dredge-up phase. 
 Chemical abundances at the end of the second dredge-up 
for the 6 and 4$M_\odot$
stars (with $Z=0.02$) are also  shown in Figures 4 and 5, respectively.

\begin{figure}
\plottwo{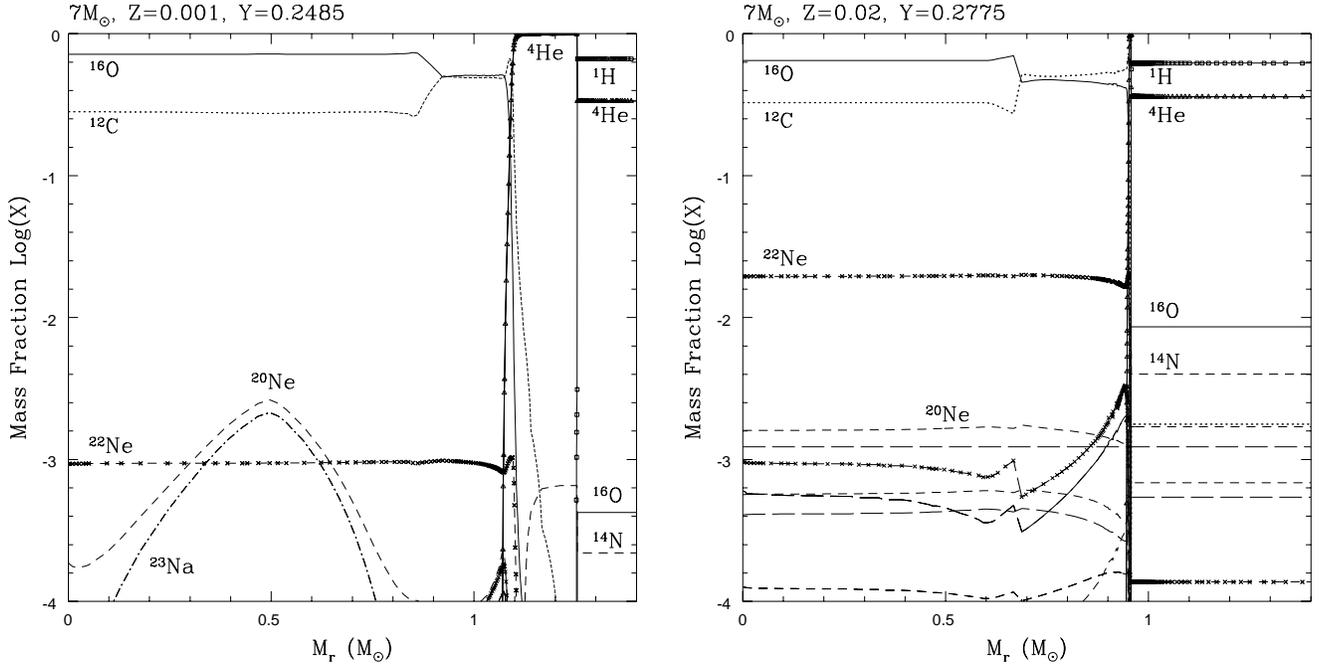}{coFIG3b.epsi}
\figcaption[coFIG3a.epsi, coFIG3b.epsi]
{(a) Abundances in mass fraction in the inner core of the 
7 $M_\odot$ 
star for  $Z$=0.001 at age $t$=48.45 Myr, with the central
temperature and density log $T_{\rm c}$ (K)=8.58 and log $\rho_{\rm
c}$ (g cm$^{-3}$) =6.82 (left panel). 
The maximum temperature, log $T_{\rm max}$=8.78, 
is attained at $M_{\rm r} = 0.54 M_\odot $.
(b) Same as Figure 3(a) but for $Z$=0.02 at the
end of the second dredge-up, $t$=51.87 Myr, with log
$T_{\rm c}$=8.29 and log $\rho_c$ =7.04 (right panel). 
Maximum temperature, log $T_{\rm max}$=8.69,  
is attained at $M_{\rm r} = 0.72 M_\odot $.\label{FIG3}}
\end{figure}

\placefigure{FIG3}

\placefigure{FIG4}
\placefigure{FIG5}

\subsection {ZAMS mass ranges of SN Ia progenitors}

 The stars that do not ignite carbon
will evolve into the asymptotic giant branch (AGB) phase. 
During this phase, 
if they are single stars, the C-O core mass increases slightly due to H-He
 double shell burning until they lose almost all hydrogen-rich
envelopes by wind mass loss. If they are
in close binary systems, their hydrogen-rich 
envelopes will be stripped off in much shorter time scale due to the
Roche lobe overflow, and C-O WDs are left.
When the companion star evolves off the main-sequence,
mass transfer occurs from the companion star to the WD. 
A certain class of these accreting WDs become SNe Ia depending on the 
accretion rate and the initial mass of the WD (e.g.,
Nomoto 1982; Nomoto \& Kondo 1991)

 The WD mass at the onset of mass transfer 
from the companion star (denoted by $M_{\rm WD,0}$)
is critically important for the fate of the
accreting WD.
For $M_{\rm WD,0}$ as low as $\lsim 0.7 M_\odot$, the WD mass is
difficult to reach $M_{\rm Ia} \simeq 1.37-1.38 M_\odot$. 
For higher  $M_{\rm WD,0}$,
the WD can become SNe Ia in a certain binary 
system  (Hachisu, Kato, \& Nomoto 1996; Li \& van den Heuvel 1996). 
However, the $M_{\rm WD,0}$ depends
on various uncertain factors, such as a binary separation,
effects of recurring He shell flashes, and a 
mass loss rate during the AGB phase.
In the following discussion, we assume that
$M_{\rm WD,0}$ is not significantly larger than
the C-O core mass at the end of the second dredge-up phase, which we denote
$M_{\rm CO}$.

  Figure 6 shows the metallicity dependence of the relation between
the ZAMS mass ($M_{\rm ms}$) and $M_{\rm CO}$. We note that 
 stars with higher metallicity form smaller C-O cores if they have
the same $M_{\rm ms}$. Our $M_{\rm ms}$ vs. $M_{\rm CO}$ relations
are similar to Castellani \etal (1990) 
which used Los Alamos Opacity (Huebner \etal 1977)
and assumed overshooting but suppressed breathing pulses at the
end of He burning. They calculated Z=0.02, 0.006, 0.002 and Y=0.23, 0.27
cases for 3, 4, 5, 7, 9 M$_\odot$ stars. 
Our $M_{\rm ms}$ vs. $M_{\rm CO}$ relations are also consistent
with  Vassiliadis \& Wood (1993), which used OPAL opacity without 
overshooting.

Figure 7 shows $M_{\rm UP}$ as a function of metallicity, which is
compared with BI79 and CCPT.
Our results in these figures are quantitatively different from BI79,
because BI79 used old  Cox \& Stewart (1970) opacity. 
Our results are also different from
CCPT's because they assumed relatively large overshooting. 
Although Castellani \etal (1990) did not estimate  $M_{\rm UP}$ exactly,
their results indicate that their  $M_{\rm UP}$ lies closer to ours
than those of  BI79 and CCPT.
CCPT and Castellani \etal (1990) assumed strong overshooting, but
we note that when OPAL opacity is used the overshooting effect 
should not be very large to meet the
observational constraints (Stothers \& Chin 1991).
Schaerer \etal (1993) also estimated  $M_{\rm UP}$ by comparing
their grids of stellar models with $T_c$ vs. $\rho_c$ tracks
in Maeder \& Meynet (1989), which is shown in Figure 7 by the crosses.
Our results show a stronger metallicity dependence
than Schaerer \etal (1993). Especially, 
their  $M_{\rm UP}$ at Z=0.001 is significantly larger than ours.

 Tornamb\'e \& Chieffi (1986) studied  $M_{\rm UP}$ for very extremely
metal-deficient stars (Z $\leq 10^{-4}$) and found that the $M_{\rm UP}$ 
increases for smaller Z. Though verifying this result may be
interesting, we do not treat these extremely low metal stars
because stars with Z $< 10^{-3}$ may not become SNe Ia progenitors
in the Chandrasekhar mass model (Kobayashi \etal 1998).

 In Figure 7, we also plot $M_{\rm LO}$, which is defined to be
the lower ZAMS mass limit for stars to become SNe Ia.
Here we assume that in order for the accreting WD to reach $M_{\rm
Ia}$, $M_{\rm CO}$ should be larger than a certain critical mass, typically
$ \sim 0.8 M_\odot$ (Hachisu \etal 1996). 
Since this critical  $M_{\rm CO}$ depends on the
binary parameters, three cases of  $M_{\rm CO}$ are shown in 
Figure 7.

\placefigure{FIG6}
\placefigure{FIG7}

In Figure \ref{FIG8} we show the number of SNe Ia progenitors, i.e. the
number of stars in the range $M_{\rm LO} \leq M \leq M_{\rm UP}$,
normalized to the case of $Z=0.02$ and $M_{\rm CO}=0.8 M_\odot$.
Since $M_{\rm LO}$ is sensitive to 
the binary parameters, we consider four cases of $M_{\rm LO}$
in Figure 8. In obtaining this figure, 
Salpter IMF is used. This figure shows that if Salpeter IMF holds also
in a lower metallicity 
environment, the number of SNe Ia progenitors increase
toward low metallicity, because stars with lower metallicity have 
larger C-O cores for the same $M_{\rm ms}$. 
It should be noted that these numbers do not necessarily 
proportional to the SNe Ia rate, because we need to take into account 
the binary companion's  lifetime in order to calculate the SNe Ia rate.

\placefigure{FIG8}

\subsection {C/O ratio in white dwarfs}

 From Figures 3-5, we find that the central part of these stars
is oxygen-rich. The C/O ratio is nearly constant in the innermost
region, which was a convective core during He burning.
Outside this homogeneous region, where the C-O layer grows due to He
shell burning, the C/O ratio 
increases up to C/O $\gsim 1$; thus the oxygen-rich core
is surrounded by a shell with 
C/O $\gsim$ 1. In fact this is a generic feature in
 all models we calculated. The C/O ratio in the shell 
is C/O $\simeq$ 1 for the star as massive as $\sim 7 M_\odot$ (Fig. 3b), 
and C/O $>1$ for less massive stars (Figs. 4 and 5). 

 In most cases of our calculations as well as previous ones, from the
center to the C-O core edge the oxygen abundance
has a peak near the outer edge of the oxygen-rich core.
This configuration is Rayleigh-Taylor unstable and will be
rehomogenized by convection (Salaris \etal 1997; Isern \etal 1998). 
Their results show that the rehomogenization proceeds slowly and
the effect is relatively small during the 
liquid phase of the WD interior. However, because of this effect
the final chemical profile of the inner SNe Ia progenitors 
may differ from the profile at the end of second dredge-up 
depending on the time scale of becoming SNe Ia.

 Figure \ref{FIG9} shows the central C/O ratio in mass fraction,
which is not a monotonic function of the metallicity and the ZAMS mass.
One reason for this complexity is that the C/O ratio is sensitive to
the evolutionary change in the
convective core size during the later stage of He burning.
Small differences in the core temperature and opacity lead to a 
different evolutionary behavior
of the convective core, which largely changes the C/O ratio. 

 The central C/O ratio is sensitive to the  $^{12}$C$(\alpha,\gamma)^{16}$O
rate. For example, for $M$ =  5$M_\odot$, $Z=0.02$, and $Y=0.27753$, 
this ratio is
0.58 for CF88$\times$1.5 rate, while   C/O = 0.89, 0.47, and 0.24
for CF88, CF88$\times$1.7, and CF85 rates, respectively.
This ratio, on the other hand, is not so sensitive to $Y$:
for example, C/O=0.58 and 0.53 for Y=0.285 and Y=0.30, respectively.
Several groups including Salaris \etal (1997) and Bressan \etal (1993)
calculated the central C \& O mass fractions for solar metallicity.
SSMM and a series of subsequent papers calculated the fractions
for various metallicities. 
Our results of the central C/O ratio is systematically larger than those
of SSMM and Salaris et al., who adopted the CF85  
$^{12}$C$(\alpha,\gamma)^{16}$O rate and different treatments of 
convection. 

\placefigure{FIG9}

Figure \ref{FIG10} shows the total  mass of  $^{12}$C ($M(^{12}$C)) 
integrated over the WD mass, $M_{\rm Ia} \simeq 1.37 M_\odot$, 
just before the SN Ia explosion.
 Here we assume C/O $\sim 1$
in the outer layer of $M_{\rm CO}\leq M_r \leq M_{\rm Ia}$ which is
formed by He shell burning during accretion. This is because our
results as well as other previous works 
indicate that a star with C-O core
mass heavier than $\sim 1 M_\odot$ tends to yield C/O $\sim 1$ in the
product of He shell burning. 
We note that C/O depends on both
the ($\rho$, $T$) track and the nuclear reaction rate. For example, if 
we use CF88 for the $^{12}$C$(\alpha,\gamma)^{16}$O rate in the model
of Figure 3(b), the C/O ratio in the shell is C/O $\simeq 1.46$.

\placefigure{FIG10}

The total carbon mass is important for studying SNe Ia, because 
it determines the total explosion energy. 
The total carbon mass fraction, $M(^{12}$C)/$M_{\rm Ia}$,
 varies in the range of 0.36 -- 0.5
depending on the ZAMS mass and metallicity, which might be related to 
the observed variation of the SNe Ia brightness.

 In Figures \ref{FIG11} and \ref{FIG12}, 
we plot the central C/O ratio and $M(^{12}$C) 
again, but abscissa is changed to $M_{\rm CO}$. These figures show that 
by replacing the abscissa from $M_{\rm ms}$ to $M_{\rm CO}$, these
curves somewhat converge. Its implication is discussed in the next 
section.

\placefigure{FIG11}
\placefigure{FIG12}

\section {Summary and Discussions}

 We have calculated the stellar evolution through the carbon ignition
or the second dredge-up for $M= 3-9 M_\odot$ and $Z= 0.001-0.03$.
Using these evolutionary models, we have made systematic
investigations on the stellar metallicity and mass dependencies
of the C-O core mass $M_{\rm CO}$ at the end of the second dredge-up,
the C/O ratio in the C-O core, and the mass range ($M_{\rm LO}$,
$M_{\rm UP}$). The major effect of changing
metallicity appears in the $M_{\rm CO} - M_{\rm ms}$ relation 
(Figs. 9 and 11).
Main conclusions and the implication on SNe Ia are
as follows:

\noindent
1) The $Z$ dependence of $M_{\rm UP}$ and  $M_{\rm LO}$
are important to obtain the white dwarf luminosity function, and to model
the galactic chemical evolution, because these determines
the SNe Ia rate.
Our results exhibit that  $M_{\rm UP}$ and $M_{\rm LO}$ are both
smaller for lower $Z$. This implies that
if the initial mass function
is independent of metallicity, the number density of 
SN Ia progenitors are larger in a lower metallicity environment
(Fig. \ref{FIG8}); therefore, the luminosity
function of WDs and SNe Ia would depend on $Z$.

\noindent
2) Our results indicate
that the innermost part of C-O WDs is likely to be
oxygen-rich (C/O $ \lsim $ 0.6) rather than C/O = 1 as usually assumed in
the SNe Ia modeling.  We find two tendencies of the central C/O ratio:
(1) For a fixed $Z$, as stellar mass
increases, the C/O ratio at the stellar center first increases to the
maximum and then decreases. The location of peaks, which depend on
metallicity, are seen at lower mass in our results (M$_{\rm ms} \sim
5 $M$_\odot$) than in SSMM (M$_{\rm ms} \sim 7 $M$_\odot$). 
(2) For larger $Z$, the above maximum
C/O ratio is larger. 
Compared with SSMM, our results show a much
larger variation in the C/O ratio distribution.

\noindent
3) For WDs which can be progenitors of SN Ia,
the central ratio varies in the range of C/O = $0.4 - 0.6$
depending mainly on the ZAMS mass.
The total carbon mass fraction just before SN Ia explosion is found to 
vary in the range $M(^{12}$C)/$M_{\rm Ia}$ = 0.36 -- 0.5.
These may be partially
responsible for the observed variations in the brightness and
spectra of SNe Ia.

 Most of previous works on the SNe Ia
explosion have used progenitor models with C/O=1. Only recently
H\"oflich \etal (1998) have started calculations with different C/O for
the DD models. 
The variation of C/O ratio would be important,
because as the inner regions are burned during the deflagration phase,
the energy release is changed and, consequently, the pre-expansion of
the outer layers is modified. This changes the total explosion energy,
and hence the expansion rate of the SNe Ia, the brightness and shape
of light curves as well.
Also it may affect the  $\rho_{tr}$ in the DD model to induce a
large variation in the explosion energy.

\noindent
4) The ranges of the variation in the central 
C/O ratio and  $M(^{12}$C)/$M_{\rm Ia}$  obtained above
do not significantly depend on
$Z$ for $0.001\leq Z\leq 0.02$. 
 These ratios shown in Figures 9 and 
10 exhibit smaller dispersion with respect to
metallicity if these ratios are plotted against $M_{\rm CO}$ 
(Figs. 11 and 12). We conclude that
the metallicity effects are relatively minor for
the same $M_{\rm CO}$, although there are
still some significant variations in the central C/O ratio for
$M_{\rm CO} \gsim 0.8M_\odot$ (Fig. 11). Moreover,
WDs with $Z \lsim 0.002$ need not be considered for SNe Ia progenitors,
because such low metallicity WDs blow too weak winds to evolve into
SNe Ia (Kobayashi \etal 1998). These would be good for 
determining cosmological parameters with the use of SNe Ia, 
because ``evolutionary'' effects
on the SN Ia properties are minor at different red-shifts. 
 
\bigskip
 We thank H. Saio for kindly providing us his code, and
P. H\"oflich, N. Prantzos, and F. -K. Thielemann for useful conversation.
This work has been supported in part by the grant-in-Aid for
Scientific Research (05242102, 06233101, 6728, 10147212) and COE research
(07CE2002) of the Ministry of Education, Science and Culture in Japan
and by the National Science Foundation in US under Grant No. PHY94-07194.

\begin{table*}[h]
\begin{center}
\caption{\centerline{Isotopes included in the nuclear network for
He burning.} \label{tab1}}
\vskip .2cm
\begin{tabular}{ccccc} \tableline\tableline
Isotope & A && Isotope & A\cr
\cline{1-2} \cline{4-5} 
\tableline
n & 1 && H & 1 \cr
He & 4 && C & 12-13 \cr
N & 13-15 && O & 14-18 \cr
F & 17-19 && Ne & 18-22 \cr
Na & 21-23 && Mg & 22-27 \cr
Al & 25-29 && Si & 27-30 \cr
Fe & 56 &&  &  \cr
\hline
\end{tabular}
\end{center}
\end{table*}

\clearpage

\begin{figure}
\plottwo{coFIG4.epsi}{coFIG5.epsi}
\figcaption[coFIG4.epsi]
{Same as Figure 3(b) but for  $M$ = 6$M_\odot$. (left) \label{FIG4}}
\figcaption[coFIG5.epsi]{
Same as Figure 3(b) but for  $M$ = 4$M_\odot$.(right) \label{FIG5}}
\end{figure}

\begin{figure}
\plotone{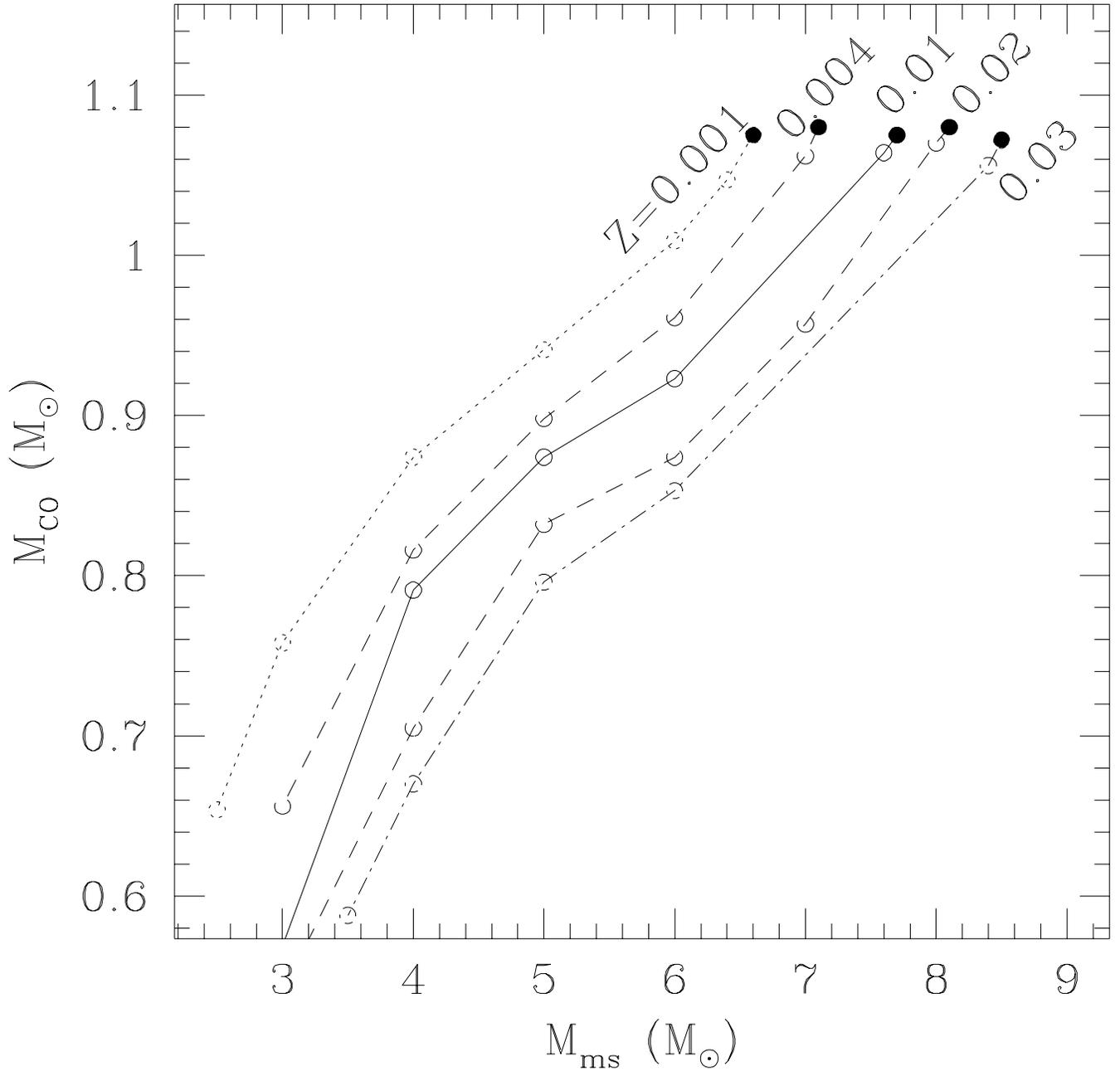}{}
\figcaption[coFIG6.epsi]
{Metallicity dependence of the ZAMS mass ($M_{\rm ms}$) vs.
C-O core mass ($M_{\rm CO}$) at the end of the second dredge-up. Filled
circles indicate that off-center carbon ignition occurs in these
model. Hence, the abscissa of these points correspond to the 
$M_{\rm UP}$.\label{FIG6} }
\end{figure}

\begin{figure}
\plotone{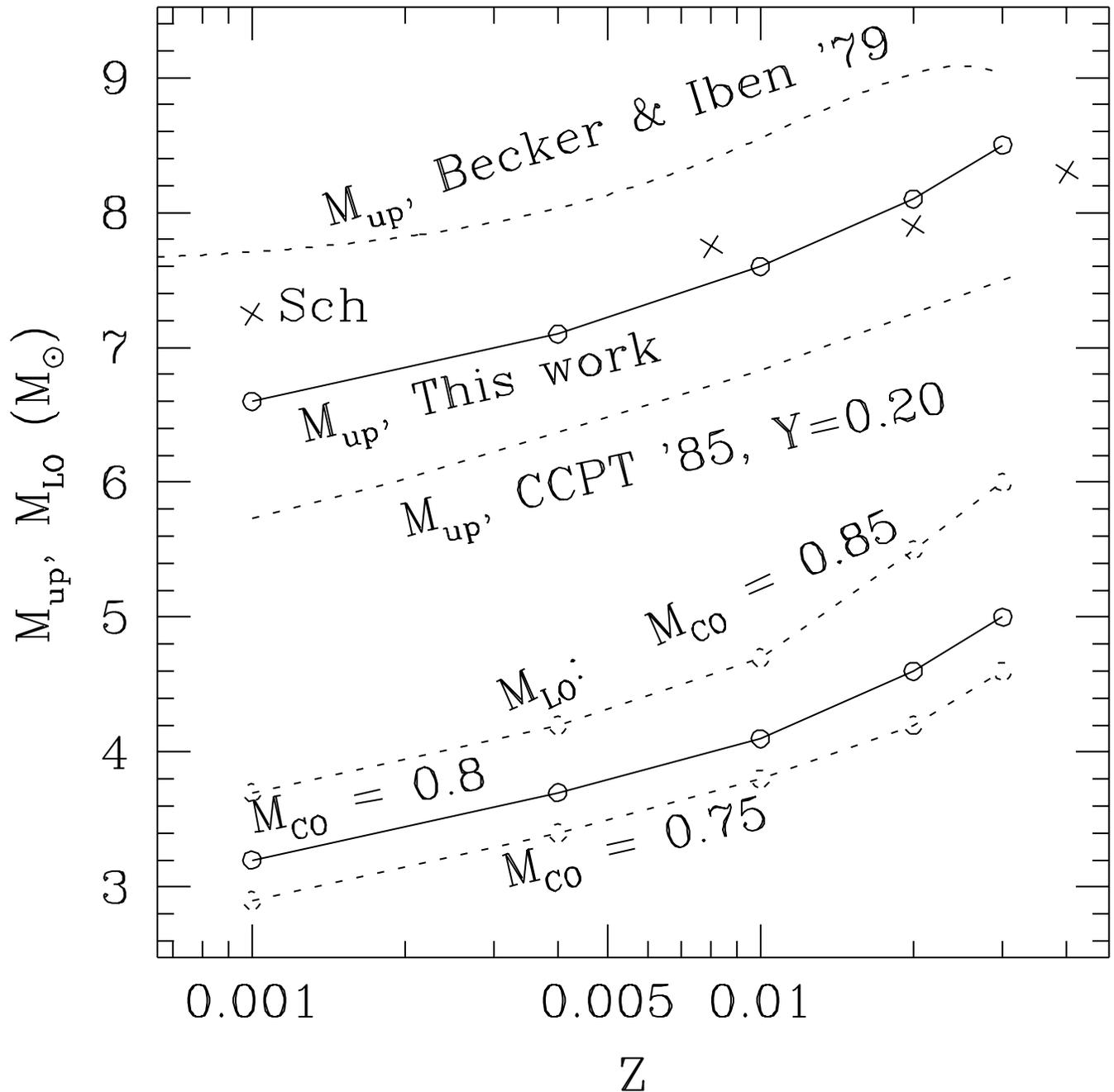}
\figcaption[coFIG7.epsi]
{ $M_{\rm UP}$ obtained in this work compared with earlier works.
The lower ZAMS mass limit for becoming an SN Ia, $M_{\rm LO}$, is also shown. 
Since this value depends on the binary parameters, three curves are shown,
which correspond to $M_{\rm CO}$ = 0.75, 0.80 and 0.85 $M_\odot$.
Crosses labeled by 'Sch' are $M_{\rm UP}$ of Schaerer \etal (1993).
\label{FIG7}}

\end{figure}

\begin{figure}
\plotone{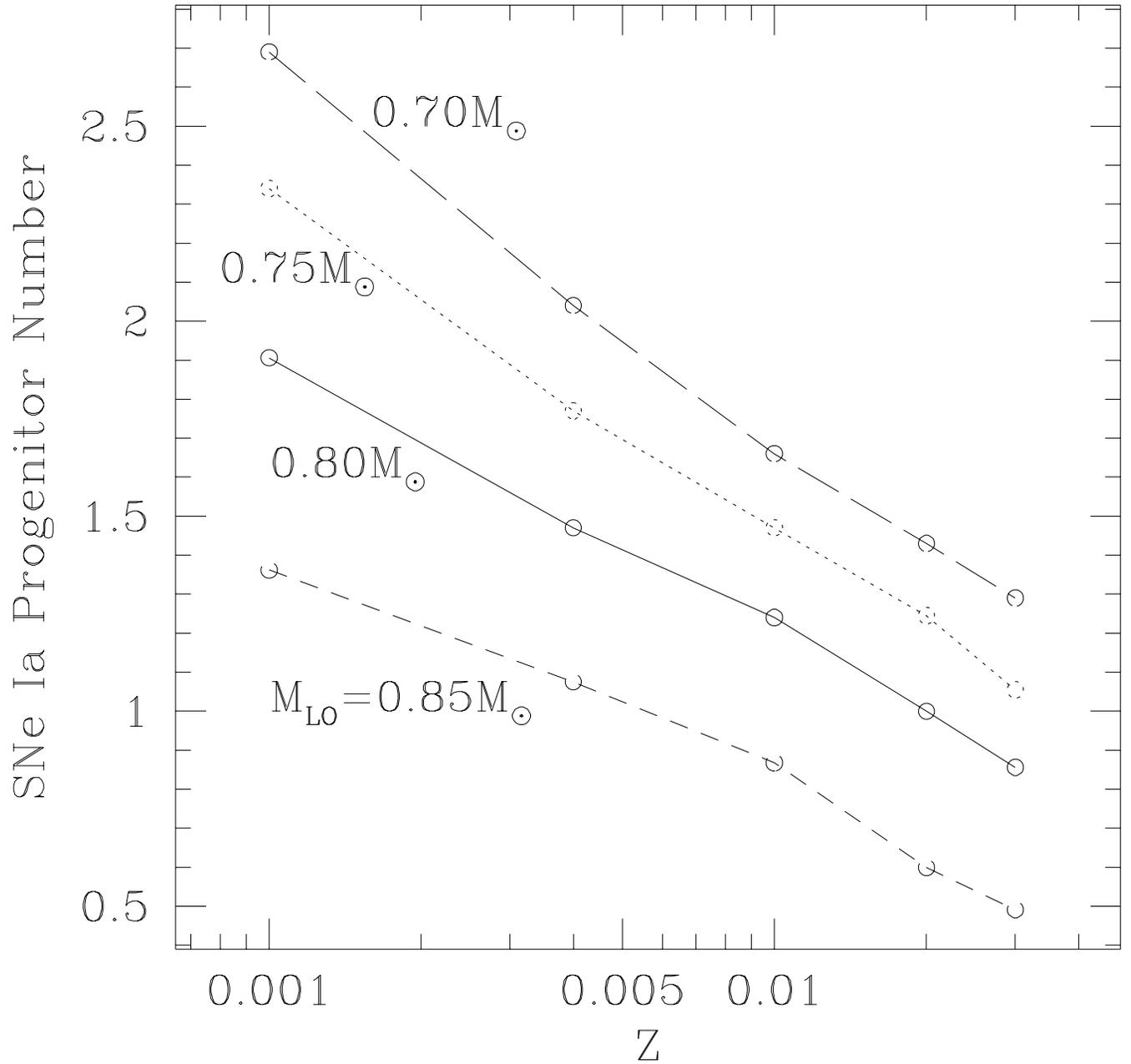}
\figcaption[coFIG8.epsi]
{Metallicity and $M_{\rm LO}$ dependence of the number of SNe Ia 
 progenitors normalized to the case with $Z=0.02$ and 
$M_{\rm LO}=0.8 M_\odot$.
The Salpeter IMF is assumed. \label{FIG8}}
\end{figure}

\begin{figure}
\plottwo{coFIG9.epsi}{coFIG10.epsi}
\figcaption[coFIG9.epsi]
{Central C/O ratio in mass fraction as a function of metallicity 
and stellar mass. (left) \label{FIG9}}
\figcaption[coFIG10.epsi]
{Total  $^{12}$C mass ($M(^{12}$C)) 
 included in the WD of mass $M_{\rm Ia} = 1.37 M_\odot$
just before an SN Ia explosion
 as a function of $M_{\rm ms}$. (right) \label{FIG10}}
\end{figure}

\begin{figure}
\plottwo{coFIG11.epsi}{coFIG12.epsi}
\figcaption[coFIG11.epsi]
{Central C/O ratio in mass fraction as a function of metallicity 
and $M_{\rm CO}$. (left) \label{FIG11}}
\figcaption[coFIG12.epsi]
{Total  $^{12}$C mass ($M(^{12}$C)) included in the
WD of mass  $M_{\rm Ia} = 1.37 M_\odot$ just before an SN Ia
explosion as a function of $M_{\rm CO}$. (right) \label{FIG12}}
\end{figure}

\end{document}